\documentstyle[preprint,aps,epsfig,equation]{revtex}
\begin{document}
\newcommand{\be}{\begin{equation}}
\newcommand{\ben}{\begin{subequations}}
\newcommand{\een}{\end{subequations}}
\newcommand{\beq}{\begin{eqalignno}}
\newcommand{\eeq}{\end{eqalignno}}
\newcommand{\ee}{\end{equation}}
\renewcommand{\thefootnote}{\fnsymbol{footnote}} 
\def\llgm{\left\lgroup\matrix}
\def\rrgm{\right\rgroup}
\draft
\title{Supersymmetric Higgs Pair Production at Hadron Colliders} 
\author{A.\ Belyaev $^{1,2}$,  
 	M.\ Drees $^{1}$, 
 	O.\ J.\ P.\ \'Eboli $^{1}$, 
	J.\ K.\ Mizukoshi $^{3}$, and 
        S.\ F.\ Novaes $^1$}
\address{$^1$ Instituto de F\'{\i}sica Te\'orica,
     	      Universidade  Estadual Paulista, \\
              Rua Pamplona 145, 01405--900, S\~ao Paulo, Brazil. \\
	 $^2$ Skobeltsin Institute for Nuclear Physics,
	      Moscow State University, 119 899, Moscow, Russia\\
         $^3$ Stanford Linear Accelerator Center, University of Stanford, \\ 
              Stanford, CA 94309, USA}
\date{\today}
\maketitle
\begin{abstract}
We study the pair production of neutral Higgs bosons through
gluon fusion at hadron colliders in the framework of the Minimal
Supersymmetric Standard Model. We present analytical expressions
for the relevant amplitudes, including both quark and squark loop
contributions, and allowing for mixing between the superpartners
of left-- and right--handed quarks. Squark loop contributions can
increase the cross section for the production of two CP--even
Higgs bosons by more than two orders of magnitude, if the
relevant trilinear soft breaking parameter is large and the mass
of the lighter squark eigenstate is not too far above its current
lower bound. In the region of large \text{$\tan \! \beta$},
neutral Higgs boson pair production might even be observable in
the $4 b$ final state during the next run of the Tevatron
collider.
\end{abstract}

\section{Introduction}

The electroweak symmetry breaking sector is the last ingredient
of the Standard Model (SM) that remains to be explored
experimentally. If the masses of the $W$ and $Z$ bosons, as well
as of the charged leptons and quarks, are due to the vacuum
expectation value of an elementary scalar Higgs field,
naturalness arguments indicate \cite{witten} that Nature should
become supersymmetric at an energy scale not far above the weak
scale; otherwise quantum corrections would destabilize the
hierarchy between the electroweak scale and the much larger
Planck, GUT or string scale. Supersymmetry not only demands the
introduction of superpartners for all know SM particles, but it
also requires the existence of at least two Higgs doublets (and
their superpartners). In its minimal version, the so called
minimal supersymmetric standard model (MSSM) \cite{mssm}, there
are three physical neutral Higgs bosons after electroweak
symmetry breaking. In this work we assume that CP is conserved in
the Higgs and squark sectors of the theory, and consequently the
neutral Higgs bosons can be classified as two CP--even states
$h,\ H$ (with $m_h < m_H$) and one CP--odd state $A$.

The production and detection of neutral Higgs bosons at $e^+e^-$
colliders is quite straightforward \cite{hunter}. It can be shown
\cite{eeh} that at least one MSSM Higgs boson would have to be
found at an $e^+e^-$ collider operating at a center--of--mass
energy $\sqrt{s} \gtrsim 250$ GeV. However, there are no immediate
prospects for the construction of such a collider, and the energy
of LEP may not suffice to find even one MSSM Higgs boson. On the
other hand, the Tevatron will soon begin its next collider run,
at a slightly increased energy ($\sqrt{s} = 2$ TeV) and greatly
increased luminosity (anticipated $\int {\cal L} dt = 2$
fb$^{-1}$ per experiment); future runs with yet higher luminosity
are being contemplated. In a few years experiments at the LHC
will also start taking data at $\sqrt{s}=14$ TeV and luminosity
between 10 and 100 fb$^{-1}/$yr. It is therefore important to
explore all channels that might yield information about the Higgs
sector at hadron colliders.

The largest contribution to the inclusive neutral Higgs
production cross section at such colliders comes from gluon
fusion, which can produce a single Higgs boson through quark
\cite{georgi,gh2} or squark \cite{gh2} loops. The
next--to--leading order (NLO) QCD corrections (involving gluons
and light quarks) to these processes have already been computed
\cite{nlo1,nlo2}. On the other hand, NLO SUSY QCD corrections
(involving gluinos) have not yet been calculated, however, they
are expected to be smaller. One drawback of this production
channel is that it can give a visible signal only if the Higgs
boson decays into a mode with a small branching ratio. For
example, in case of the light CP--even scalar $h$, the only
promising decay mode in this production channel is $h \rightarrow
\gamma \gamma$, which has a branching ratio of the order of $
10^{-3}$ \cite{gamberini}. Moreover, if squark loop contributions
are large, they tend to be destructive \cite{stoploop}, at least
for $gg \rightarrow h \to \gamma \gamma$. As a result, $gg
\rightarrow \phi$ production in the MSSM ($\phi = h, H, A$)
cannot give a viable signal at the Tevatron, and is often
difficult to detect even at the LHC.

One alternative is to search for the production of a single Higgs
boson produced in association with some other particle(s)
\cite{assoc}. However, the main (hadronic) decay channels of MSSM
Higgs bosons seem to be detectable in this way only in $W h$ ($Z
h$) production at the Tevatron \cite{wh} if more than 10
fb$^{-1}$ of data can be accumulated, as well as in some cases
for $h$ bosons produced in the cascade decays of heavy sparticles
at the LHC \cite{hcascade}. The detectability of hadronic Higgs
decays in $t \bar{t} \phi$ production \cite{tth} still awaits
confirmation by an experimental study; no background estimates
for the recently suggested \cite{ststh} $\tilde{t} \tilde{t}
\phi$ channels exists as yet. Finally, for very large values of
\text{$\tan \! \beta$}\ associate Higgs $b \bar{b}$ production
might also be observable at the Tevatron in the $4b$ \cite{yuan}
and/or the $b \bar{b} \tau^+ \tau^-$ \cite{mono} mode.

In this paper we instead study the production of {\em two}
neutral Higgs bosons. It is hoped that the $4b$ final state, with
invariant mass peaks in both $b \bar{b}$ pairs, will give a
detectable signal at the Tevatron and/or the LHC at least in some
regions of parameter space. Moreover, the Higgs--boson pair
production can in principle be used to probe the Higgs
self-couplings in order to reconstruct its potential \cite{dkmz}.
This process has first been discussed in refs.~\cite{glover} for
the SM, and in \cite{plehn} the quark loop contribution in the
MSSM has been studied. NLO QCD corrections to heavy quark loops
have also recently been calculated \cite{nlo3}; they are of
similar relative size as the corresponding NLO corrections to
single Higgs production. Here we extend these analyses by
including the contribution from squark loops, allowing for
general mixing of the superpartners of left-- and right--handed
quarks. We find that, unlike for single Higgs production
\cite{stoploop}, squark loop contributions can increase the total
cross section by more than two orders of magnitude. In some
regions of parameter space beyond the reach of LEP, the $4b$
final state might even give a detectable MSSM Higgs signal at the
next run of the Tevatron.

The primary purpose of this paper is to present all analytical
expressions required for a calculation of the squark loop
contribution to the production of two neutral Higgs bosons, and
to illustrate their potential importance with a few numerical
examples. 

The remainder of this paper is organized as follows. In Sec.~II we
give analytical expressions for all one--loop amplitudes of the
form $g g \rightarrow \phi_i \phi_j$, where $\phi_{i,j}$ are
neutral Higgs bosons with definite CP quantum numbers. These
expressions are completely general; a list of the relevant
couplings as predicted by the MSSM is given in the Appendix. In
Sec.~III some numerical results are shown. We focus on scenarios
with either a small or a large value of the parameter \text{$\tan
\! \beta$}, which governs the size of Yukawa couplings in the
MSSM. We find potentially very large squark loop contributions in
both cases, but the experimental discovery of Higgs boson pairs
at the Tevatron seems to be somewhat more promising at large
\text{$\tan \! \beta$}. Finally, Sec.~IV is devoted to a brief
summary and some conclusions.

\section{Cross Sections and Matrix Elements}

We write the subprocess (parton--parton) differential cross section as,
\begin{equation} \label{e1}
{d\hat{\sigma} \over d\hat{t}}={|M|^2 \over 16\pi \hat{s}^2} \;\; ,
\end{equation}
where the hatted Mandelstam variables refer to the parton--parton
system. The total cross section can be obtained from (\ref{e1})
by integrating over $\hat{t}$ and convoluting with the gluon
densities in the two colliding hadrons, as usual. In general, the
invariant amplitude can be written as
\ben \label{e2} \beq
|M|^2 = 2 \left({1\over 4}\right) \left( {1\over 64} \right) S_F
\Bigl[
  \left|\sum_n M_{++}^{(n)} (\phi_i \phi_j)\right|^2
&+ \left|\sum_n M_{--}^{(n)} (\phi_i \phi_j) \right|^2 \nonumber \\
 + \left|\sum_n M_{+-}^{(n)} (\phi_i \phi_j)\right|^2
&+ \left|\sum_n M_{-+}^{(n)} (\phi_i \phi_j)\right|^2
\Bigr] \;\; .
\eeq \een

Here $M_{\lambda_1, \lambda_2}^{(n)} (\phi_i \phi_j)$ is the
helicity amplitude for the production of the Higgs boson pair
$(\phi_i \phi_j)$ for the initial gluon helicities
$\lambda_{1(2)} = \pm$. The sum runs over all Feynman diagrams
$(n)$ that contribute to a specific process. The factors refer
to the average over the initial gluon helicities ($\frac{1}{4}$),
the color factor $[\rm{Tr}(T^a T^b)]^2 = \frac{1}{4} \delta_{aa}
= 2$, and the average over the gluon colors ($\frac{1}{64}$). The
symmetry factor, $S_F$, equals $\frac{1}{2}$ for the production
of two identical Higgs bosons, and is $1$ otherwise. 

\begin{figure}
\mbox{\psfig{file=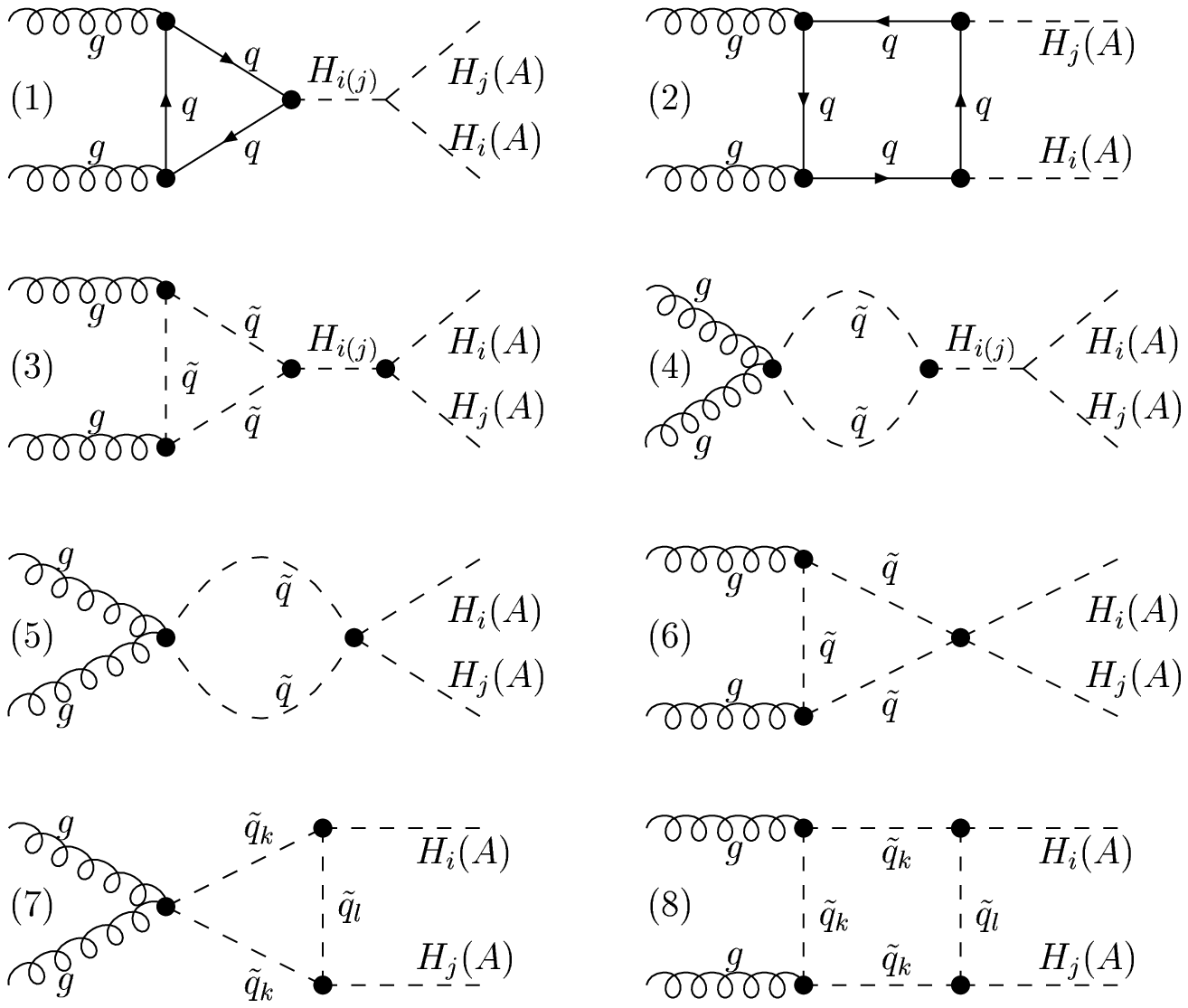,width=\textwidth}}
\protect
\caption{Feynman diagrams for  $hh$, $HH$, $hH$, and $AA$ Higgs boson pair
production. $H_{i(j)} = h, H$ for $i(j)=1,2$ respectively, $\tilde
q_{k(l)} = \tilde q_1, \tilde q_2$ for $k(l)=1,2$. The crossed
diagrams are not shown.}
\label{fig:hh}
\end{figure}

\begin{figure}
\mbox{\epsfig{file=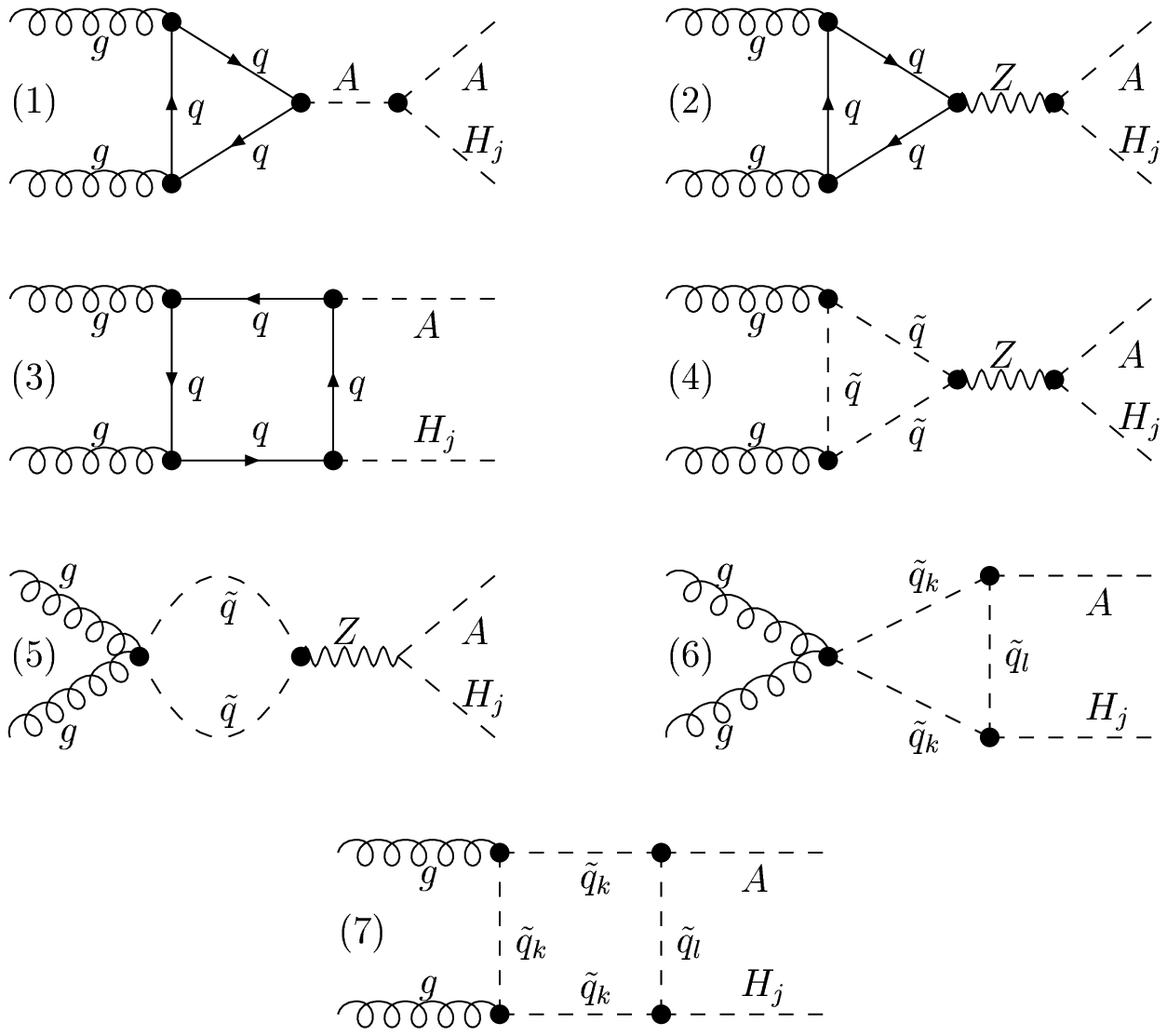,width=\textwidth}}
\protect
\caption{Feynman diagrams for the $hA$ and $HA$ Higgs boson
pair production. $H_j = h, H$ for $j=1,2$ respectively, $\tilde
q_{k(l)} = \tilde q_1, \tilde q_2$ for $k(l) = 1,2$. The crossed
diagrams are not shown.}
\label{fig:ah}
\end{figure}

The Feynman diagrams contributing to the $g g \to hh$, $HH$,
$hH$, and $AA$ processes are presented in Fig.\ \ref{fig:hh},
while the contributions to the processes $g g \to hA$ and $HA$
are shown in Fig.\ \ref{fig:ah}.  We now list the resulting
helicity amplitudes for these two classes of processes, starting
with the case where both produced Higgs boson have the same CP
properties.  We have used the FeynCalc package \cite{feyncalc}
for the analytical calculation.

\subsection{Invariant Amplitudes for $g g \to hh$, $HH$, $hH$, and $AA$}

\subsubsection{Quark Loop Contributions}

By assumption these processes conserve CP, which implies that
$M_{++} = M_{--}$ and $M_{+-} = M_{-+}$.  The two independent
helicity amplitudes for the production of two CP--even Higgs
bosons $H_i H_j$ ($i,j=1,2$ for $h$ and $H$ respectively), where
the superscript on the amplitudes refers to the number of the
Feynman diagram in Fig.~1, are as follows. 
\ben \label{e3} \beq
M_{++}^{(1)}(H_iH_j) &=
\frac {-i \alpha_s m_q V_{H_k qq} V_{H_i H_j H_k}} 
{\pi[(\hat{s}-{m_{H_k}}^2) + i m_{H_k}
\Gamma_{H_k}] }  \left[ 2 + \left( 4 {m_q}^2
- \hat{s}) C(0,0,\hat{s} \right) \right] \;\; ;
\label{e3a} \\ 
M_{+-}^{(1)}(H_i H_j) &= 0 \;\; ;
\label{e3b} \\
M_{++}^{(2)}(H_iH_j) &= 
{i\alpha_s\over{2\pi\hat{s}}}  V_{H_iqq}  V_{H_jqq} \left\{
-4 \hat{s}-8 m_q^2 C(0,0,\hat{s})\hat{s} - 
\left(8 m_q^2-m_{H_i}^2-m_{H_j}^2 \right)
\left[ T_i C(m_{H_i}^2, 0, \hat{t}) 
\right.\right. \nonumber \\ &\hspace*{5mm} 
+ \left.\left. U_i C(m_{H_i}^2,0,\hat{u}) +  T_j C(m_{H_j}^2, 0, \hat{t}) 
+ U_j C(m_{H_j}^2,0,\hat{u}) 
\right.\right. \nonumber \\ &\hspace*{5mm}
- \left.\left. (m_{H_j}^2m_{H_j}^2-\hat{t}\hat{u})
  D(m_{H_i}^2,0,m_{H_j}^2,0,\hat{t},\hat{u})\right] 
\right. \nonumber \\ &\hspace*{5mm}
- \left. 2 m_q^2 (8
m_q^2-m_{H_i}^2-m_{H_j}^2-\hat{s})\hat{s}\right.\times
\\ & \hspace*{10mm}\left.\left[ 
 D(m_{H_i}^2,0,m_{H_j}^2,0,\hat{t},\hat{u})
+D(m_{H_i}^2, m_{H_j}^2, 0, 0, \hat{s}, \hat{t})
+D(m_{H_i}^2, m_{H_j}^2, 0, 0, \hat{s}, \hat{u})\right]\right\} \;\; ;
\label{e3c} \nonumber \\
M_{+-}^{(2)}(H_i H_j) &=
\frac {-i\alpha_s V_{H_iqq}  V_{H_jqq}} 
{2\pi (m_{H_i}^2 m_{H_j}^2-\hat{t}\hat{u})}  
\left\{ 
 (8 m_q^2 - \hat{t} - \hat{u}) (2 m_{H_i}^2 m_{H_j}^2 - \hat{t}^2 - \hat{u}^2)
  C(m_{H_i}^2, m_{H_j}^2, \hat{s}) 
\right. \nonumber \\ &\hspace*{5mm}
+  (m_{H_i}^2m_{H_j}^2-8m_q^2\hat{t}+\hat{t}^2)\times
\nonumber  \\ &\hspace*{10mm}
\left[ T_i C(m_{H_i}^2, 0, \hat{t})+ T_j C(m_{H_j}^2, 0,
\hat{t}) - \hat{s} C(0,0,\hat{s})+ 
\hat{s}\hat{t} D(m_{H_i}^2, m_{H_j}^2, 0, 0, \hat{s}, \hat{t})\right]
\nonumber \\ &\hspace*{5mm}
+ (m_{H_i}^2 m_{H_j}^2 - 8 m_q^2 \hat{u} + \hat{u}^2)\times
\nonumber \\ &\hspace*{10mm}
 \left[ U_i  C(m_{H_i}^2, 0, \hat{u})+ U_j C(m_{H_j}^2, 0,
\hat{u}) - \hat{s} C(0,0,\hat{s})+ 
\hat{s}\hat{u} D(m_{H_i}^2,  m_{H_j}^2, 0, 0, \hat{s},
\hat{u})\right]  
\nonumber \\ &\hspace*{5mm} 
+ \left. 2 m_q^2 (m_{H_i}^2 m_{H_j}^2-\hat{t}\hat{u})
 (8 m_q^2 - \hat{t} -\hat{u} ) \times
\right. \\ &\hspace*{10mm} \left.
\left[ D(m_{H_i}^2, 0, m_{H_j}^2, 0, \hat{t}, \hat{u})+D(m_{H_i}^2,
m_{H_j}^2, 0, 0, \hat{s}, \hat{t})+D(m_{H_i}^2, m_{H_j}^2, 0, 0,
\hat{s}, \hat{u}) \right] \right\} \;\; .
\label{e3d}
\eeq \een
Here, $V_{H_k qq}$ and $V_{H_i H_j H_k}$ are the Yukawa coupling
constants of $H_k$ to quarks and the trilinear Higgs couplings,
respectively. Expressions for these couplings in the MSSM are given in
Appendix A while we list in Appendix B our choice for the polarization
vectors.  We have also defined the quantities
\be \label{e5}
T_i = (m_{H_i}^2 -\hat{t}) \;\; \mbox{and}  \;\; U_i = (m_{H_i}^2
-\hat{u})  \;\;  .
\ee
Furthermore, the loop integrals appearing in Eqs.~(\ref{e3}) are
defined in terms of the Passarino--Veltman scalar functions $C_0$ and
$D_0$ \cite{passa} (see Appendix B for our conventions) as
\ben \label{e6} \beq
C(a,b,c) &= C_0(a,b,c, m_q, m_q, m_q) \;\; ;
\label{e6a} \\
D(a,b,c,x,y,z) &= D_0(a,b,c,x,y,z,m_q,m_q,m_q,m_q) \;\; .
\label{e6b}
\eeq \een

The scattering amplitudes for the production of a pair of
pseudo--scalar Higgs boson are
\ben \label{e4} \beq
M_{++}^{(1)}(AA) & = M_{++}^{(1)}(H_iH_j) 
[V_{H_iH_j H_j} \to V_{H_iAA} ] \;\; ;
\label{e4a} \\
M_{+-}^{(1)} (AA) &= 0 \;\; ;
\label{e4b} \\
M_{++}^{(2)}(AA) &= 
{i\alpha_s\over{\pi\hat{s}}} V_{Aqq}  
V_{Aqq}
  \left\{ 2 \hat{s}+4 m_q^2 C(0,0,\hat{s})\hat{s}
\right. \nonumber \\ &\hspace*{5mm}
- 2 m_A^2 \left[ T_A C(m_{A}^2, 0, \hat{t}) + U_A  C(m_{A}^2, 0,
\hat{u}) \right] + m_A^2 (m_A^4 - \hat{u}\hat{t}) D(m_A^2, 0, m_A^2,
0, \hat{t}, \hat{u})
\nonumber \\ &\hspace*{5mm}
-  m_q^2\hat{s}(\hat{t}+\hat{u}) \times
\\ &\hspace*{10mm} 
\left. \left[
D(m_A^2, 0, m_A^2, 0, \hat{t}, \hat{u}) + 
D(m_A^2, m_A^2, 0, 0, \hat{s}, \hat{t}) + 
D(m_A^2, m_A^2, 0, 0, \hat{s}, \hat{u})\right] \right\} \;\; ;
\label{e4c} \\
M_{+-}^{(2)}(AA) &= {-i\alpha_s\over{2\pi (m_{A}^4 -\hat{t}\hat{u})}} 
V_{Aqq} V_{Aqq} 
\left\{
  \hat{{s}}(2m_A^4 + \hat{t}^2 + \hat{{u}}^2) C(0,0,\hat{s}) - 
  2 T_A(m_A^4 + \hat{t}^2) C(m_{A}^2, 0, \hat{t})
\nonumber  \right. \\ &\hspace*{5mm}
- \left. 
  2 U_A (m_A^4 + \hat{u}^2)  C(m_{A}^2, 0, \hat{u}) 
+ (\hat{t} + \hat{u}) (2 m_A^4 - \hat{t}^2 - 
\hat{u}^2) C(m_{A}^2, m_{A}^2, \hat{s})
\nonumber  \right. \\ &\hspace*{5mm} \left.
+  2 m_q^2 (\hat{t} + \hat{u}) (m_A^4 - \hat{t} \hat{u}) \times
\right. \\ &\hspace*{10mm}
\left[ D(m_{A}^2, 0, m_{A}^2, 0, \hat{t}, \hat{u}) + 
 D(m_{A}^2, m_{A}^2, 0, 0, \hat{s}, \hat{t}) + 
 D(m_{A}^2, m_{A}^2, 0, 0, \hat{s}, \hat{u})\right] 
\nonumber \\ &\hspace*{5mm}
- \left.\hat{s}\left[ \hat{t}(m_A^4+\hat{t}^2)
 D(m_{A}^2, m_{A}^2, 0, 0, \hat{s}, \hat{t}) +
 \hat{u}(m_A^4+\hat{u}^2)D(m_{A}^2, m_{A}^2, 0, 0, \hat{s}, \hat{u})\right]
 \right\} \;\; .
\label{e4d}
\eeq \een
Analogously to the production of CP--even Higgs bosons, we have defined 
the quantities
\be \label{e5b}
T_A = (m_{A}^2 -\hat{t}) \;\;  \mbox{and} \;\; U_A = (m_{A}^2 -\hat{u}) \;\; .
\ee
As a check of our calculations we verified that our results for
the quark (squark) loop contributions to the Higgs pair
production are invariant under QCD gauge transformations.
Furthermore, our results agree with those of ref.~\cite{plehn}.

\subsubsection{Squark Loop Contributions}

We now turn to the new results of this paper, {\em i.e.} the
squark loop contributions depicted in diagrams (3)--(8) in
Fig.~1. These can be grouped into three sets of diagrams,
$(3)+(4)$, $(5)+(6)$ and $(7)+(8)$, which are gauge invariant and
finite. Moreover, diagrams (7) and (8) are finite by themselves
and we therefore list their contributions separately, treating
QCD interactions in the 't Hooft--Feynman gauge.
\ben \label{e7} \beq
M_{++}^{(3+4)}(H_iH_j)
&= \frac {i\alpha_s V_{H_l \tilde{q}_k \tilde{q}_k }  V_{H_lH_iH_j}} 
{ 2\pi[(\hat{s}-m_{H_l}^2) + i m_{H_l} 
\Gamma_{H_l} ] }
\left[ 1 + 2 m_{\tilde{q}_k}^2 C_{kkk}(0,0,\hat{s}) \right] \;\; ;
\label{e7a} \\
M_{+-}^{(3+4)} (H_iH_j) &= 0 \;\; ;
\label{e7b} \\
M_{++}^{(5+6)}(H_iH_j) & =
{-i\alpha_s\over{2\pi}} V_{H_iH_j\tilde{q}_k \tilde{q}_k}
\left[ 1 + 2 m_{\tilde{q}_k}^2 C_{kkk}(0,0,\hat{s}) \right] \;\; ;
\label{e7c} \\
M_{+-}^{(5+6)}(H_iH_j) &= 0 \;\; ;
\label{e7d} \\
M_{++}^{(7)}(H_iH_j) &= 
{-i\alpha_s\over{\pi}}  V_{H_i\tilde{q}_k \tilde{q}_l}
 V_{H_j \tilde{q}_k \tilde{q}_l}  C_{klk}(m_{H_i}^2, m_{H_j}^2,\hat{s})
\;\; ; 
\label{e7e} \\
M_{+-}^{(7)}(H_iH_j) &= 0  \;\; ;
\label{e7f} \\
M_{++}^{(8)}(H_iH_j) & =
   {i\alpha_s\over{2\pi\hat{s}}}
   V_{H_i\tilde{q}_k\tilde{q}_l} V_{H_j \tilde{q}_k \tilde{q}_l } 
\nonumber \\
&\times \Bigl\{ T_i C_{lkk}(m_{H_i}^2,0,\hat{t})  +
U_i C_{kll}(m_{H_i}^2,0,\hat{u}) + 
T_j C_{kll}(m_{H_j}^2,0,\hat{t}) 
\nonumber \\ &\hspace*{5mm}
+ U_j C_{lkk}(m_{H_j}^2,0,\hat{u}) 
+ 2\hat{s}  C_{klk}(m_{H_i}^2,m_{H_j}^2,\hat{s}) +
\nonumber \\ &\hspace*{5mm}
\left[( m_{\tilde{q}_l}^2 -m_{\tilde{q}_k}^2 ) \hat{s} 
- ( m_{H_i}^2 m_{H_j}^2 -\hat{t} \hat{u} )
\right] D_{lkkl}(m_{H_i}^2,0,m_{H_j}^2,0,\hat{t},\hat{u}) 
\nonumber \\ &\hspace*{5mm}
+ 2\hat{s} m_{\tilde{q}_k}^2 \left[ 
 D_{lkkl}(m_{H_i}^2,0,m_{H_j}^2,0,\hat{t},\hat{u})
+D_{klkk}(m_{H_i}^2,m_{H_j}^2,0,0,\hat{s},\hat{t})
\nonumber \right. \\ &\hspace*{20mm} \left.
+D_{klkk}(m_{H_i}^2,m_{H_j}^2,0,0,\hat{s},\hat{u})\right] \Bigr\} \;\; ;
\label{e7g} \\
M_{+-}^{(8)}(H_iH_j) &= 
   {i\alpha_s\over{2 \pi(m_{H_i}^2 m_{H_j}^2-\hat{t}\hat{u})}}
   V_{H_i \tilde{q}_k \tilde{q}_l} V_{H_j \tilde{q}_k \tilde{q}_l }
\nonumber \\ &
 \times \Bigl\{ \hat{s}(2 m_{\tilde{q}_k}^2 - 2 m_{\tilde{q}_l}^2 + \hat{t}
   + \hat{u}) C_{kkk}(0,0,\hat{s})
\nonumber  \\ &\hspace*{5mm}
- \hat{t} \left[ T_i C_{lkk}( m_{H_i}^2, 0, \hat{t} ) + T_j 
C_{lkk}( m_{H_j}^2, 0, \hat{t} ) \right] 
\nonumber  \\ &\hspace*{5mm}
- \hat{u} \left[ U_i C_{lkk}(m_{H_i}^2, 0, \hat{u} ) +
U_j C_{lkk}(m_{H_j}^2, 0, \hat{u}) \right]
\nonumber \\ &\hspace*{5mm}
- T_j ( m_{\tilde{q}_k}^2 - m_{\tilde{q}_l}^2 )
\left[ C_{kll}(m_{H_j}^2, 0, \hat{t}) +
C_{lkk}(m_{H_j}^2, 0, \hat{t}) \right]
\nonumber \\ &\hspace*{5mm} 
- U_i(m_{\tilde{q}_k}^2 - m_{\tilde{q}_l}^2)
\left[C_{kll}(m_{H_i}^2, 0, \hat{u}) + C_{lkk}(m_{H_i}^2, 0, \hat{u})
\right]  
\nonumber \\ &\hspace*{5mm}
+\left( 2 m_{H_i}^2 m_{H_j}^2 - \hat{t}^2 - \hat{u}^2 \right) 
C_{klk}(m_{H_i}^2, m_{H_j}^2, \hat{s})
\\ &\hspace*{5mm}
+ \left[ - \hat{s} \left( m_{\tilde{q}_k}^2 - m_{\tilde{q}_l}^2
   \right)^2 + \left( m_{\tilde{q}_k}^2 + m_{\tilde{q}_l}^2 \right)
 \left( m_{H_i}^2 m_{H_j}^2 - \hat{t} \hat{u} \right) \right]
\nonumber \\ &\hspace*{10mm}
\times \left[
D_{lkkl}(m_{H_i}^2, 0, m_{H_j}^2, 0,\hat{t}, \hat{u}) +
D_{klkk}(m_{H_i}^2, m_{H_j}^2, 0, 0, \hat{s}, \hat{t})
\right. \nonumber \\ &\hspace*{15mm} \left.
+D_{klkk}(m_{H_i}^2, m_{H_j}^2, 0, 0, \hat{s}, \hat{u})
\right]
\nonumber \\ &\hspace*{5mm}
+ \left[ -\hat{s} \hat{t}^2 -
\left( m_{\tilde{q}_k}^2 - m_{\tilde{q}_l}^2 \right)
\left( 2\hat{t} \hat{s} - ( m_{H_i}^2 m_{H_j}^2 - \hat{t} \hat{u}) 
\right) \right]
D_{klkk}(m_{H_i}^2, m_{H_j}^2, 0, 0, \hat{s}, \hat{t})
\nonumber \\ &\hspace*{5mm}
+ \left[ -\hat{s} \hat{u}^2 - 
\left( m_{\tilde{q}_k}^2 - m_{\tilde{q}_l}^2 \right)
\left( 2 \hat{u} \hat{s} - (m_{H_i}^2 m_{H_j}^2 - \hat{t} \hat{u}) 
\right) \right]
D_{klkk}(m_{H_i}^2, m_{H_j}^2, 0, 0, \hat{s}, \hat{u}) 
\Bigr\} \;\; .
\label{e7h}
\nonumber
\eeq \een
MSSM predictions for the trilinear Higgs--squark--squark couplings
$V_{H_l \tilde{q}_k \tilde{q}_k}$ and the quartic
Higgs--Higgs--squark--squark couplings $V_{H_i H_j \tilde{q}_k
\tilde{q}_k}$ are given in Appendix A; note that we need the former
two couplings only for two identical squarks. The loop functions
appearing in Eqs.~(\ref{e7}) depend on the squark masses and are 
defined as
\ben \label{e9} \beq
C_{ijk}(a,b,c) &=
C_0(a,b,c, m_{\tilde{q}_i}, m_{\tilde{q}_j}, m_{\tilde{q}_k}) \;\; ; 
\label{e9a} \\ 
D_{ijkl}(a,b,c,x,y,z) &= 
D_0(a,b,c,x,y,z, m_{\tilde{q}_i}, m_{\tilde{q}_j}, m_{\tilde{q}_k},
m_{\tilde{q}_l}) \;\; .
\label{e9b}
\eeq \een

The corresponding expressions for a pair of pseudo--scalar Higgs
bosons are:
\ben \label{e8} \beq
M_{++}^{(3+4)}(AA) & = M_{++}^{(3+4)}(H_iH_j) 
[ V_{H_lH_iH_j} \to V_{H_lAA} ] \;\; ;
\label{e8a} \\
M_{+-}^{(3+4)}(AA) & = 0 \;\; ;
\label{e8b} \\
M_{++}^{(5+6)}(AA) & = M_{++}^{(5+6)}(H_iH_j)  
[V_{H_iH_j \tilde{q}_k \tilde{q}_k} \to V_{AA \tilde{q}_k \tilde{q}_k
}] \;\; ;
\label{e8c} \\
M_{+-}^{(5+6)}(AA) &=  0 \;\; ;
\label{e8d} \\
M_{++}^{(7)}(AA) &= M_{++}^{(7)}(H_iH_j) 
[ V_{H_{(i,j)} \tilde{q}_k \tilde{q}_k } \to V_{A \tilde{q}_k
\tilde{q}_l} ] \;\; ;
\label{e8e} \\
M_{+-}^{(7)}(AA) &= 0 \;\; ;
\label{e8f} \\
M_{++}^{(8)}(AA) &= - M_{++}^{(8)}(H_iH_j) 
[ V_{H_{(i,j)} \tilde{q}_k \tilde{q}_l }\to V_{A \tilde{q}_k
\tilde{q}_l } ] \;\; ;
\label{e8g} \\
M_{+-}^{(8)}(AA) &= -M_{+-}^{(8)}(H_iH_j) 
[ V_{H_{(i,j)} \tilde{q}_k \tilde{q}_l } \to V_{A \tilde{q}_k
\tilde{q}_l }] \;\; .
\label{e8h}
\eeq \een

\subsection{Invariant Amplitudes for $g g \to Ah$, $AH$}

We now turn to the production of two Higgs bosons with different
CP quantum numbers which only receive contributions from quark
loops. Since we assumed CP--invariance in the Higgs and squark
sectors, the squark contributions to the production of a CP--even
and a CP--odd Higgs boson, {\it i.e.} the diagrams (4), (5), (6),
and (7) of Fig.\ \ref{fig:ah} add to zero. Let us have a closer
look at this. Note that $A$ only couples to two {\em different}
squark mass eigenstates; this immediately eliminates the
equivalent of diagrams (5) and (6) in Fig.~1. Moreover, $V_{A
\tilde{q}_1 \tilde{q}_2} = - V_{A \tilde{q}_2 \tilde{q}_1}$, and
consequently the two possible orientations of the loop momentum
in diagrams (6) and (7) in Fig.~2 exactly cancel each other.
Finally a same--flavor $\tilde{q} \tilde{q}^*$ pair in a
color--singlet state coupling to a $Z$ boson is in a CP--even
state, while one CP--even and one CP--odd Higgs boson coupling to
a $Z$ boson are in a CP--odd state; this explains why diagrams
(4) and (5) vanish. For completeness we give expressions for the
quark loop contribution in our notation; our results agree with
those of ref.~\cite{plehn}. Note that we again only have two
independent helicity amplitudes since $M_{++}=-M_{--}$ and
$M_{+-}=-M_{-+}$ for $Ah$ and $AH$ production.
\ben \label{e10} \beq
M_{++}^{(1)}(AH_i) &= 
{{i\alpha_s m_q}\over{\pi[(\hat{s}-{m_A}^2) + i m_A \Gamma_A]}} 
V_{Aqq} V_{H_j AA}~ \hat{s}~ C(0,0,\hat{s}) \;\; ;
\label{e10a} \\
M_{+-}^{(1)}(AH_i) &= 0 \;\; ;
\label{e10b} \\
M_{++}^{(2)}(AH_i) &= 
{{-i\alpha_s g_Z T_{3}^q (m_{H_i}^2-m_A^2)(\hat{s}-M_Z^2)}
\over{2\pi  m_Z^2[(\hat{s}-{m_Z}^2)+im_Z\Gamma_Z]}} V_{ZAH_i} 
\left[ 1 + 2m_q^2 C(0,0,\hat{s})  \right] \;\; ;
\label{e10c} \\
M_{+-}^{(2)}(AH_i) &= 0 \;\; ;
\label{e10d} \\
M_{++}^{(3)}(AH_i) &= 
{i\alpha_s\over{2\pi\hat{s}}} V_{H_iqq}  V_{Aqq}
\nonumber \\
& \times \Bigl\{(m_A^2 - m_{H_i}^2) \Bigl[
T_A C(m_{A}^2,0,\hat{t})+ T_i C(m_{H_i}^2,0,\hat{t}) + 
U_A C(m_{A}^2,0,\hat{u})
\nonumber \\ &\hspace*{5mm}
+ U_i C(m_{H_i}^2,0,\hat{u}) - (m_A^2 m_{H_i}^2-\hat{t}\hat{u})
D(m_{A}^2,0,m_{H_i}^2,0,\hat{t},\hat{u}) \Bigr]
\nonumber \\ &\hspace*{5mm}
+ 2 m_q^2 \hat{s} (T_A+U_A)\left[ 
D(m_{A}^2,0,m_{H_i}^2,0,\hat{t},\hat{u}) +
 D(m_{A}^2,m_{H_i}^2,0,0,\hat{s},\hat{t})
\right. \nonumber \\ &\hspace*{35mm} \left.
+  D(m_{A}^2,m_{H_i}^2,0,0,\hat{s},\hat{u})
\right]
\Bigl\} \;\; ;
\label{e10e} \\
M_{+-}^{(3)}(AH_i) &= 
{-i\alpha_s\over{2\pi (m_{A}^2 m_{H_i}^2-\hat{t}\hat{u})}} 
V_{H_iqq}  V_{Aqq}
\nonumber \\ &
\times \Bigl\{ \hat{s} (\hat{t}^2 - \hat{u}^2) C(0,0,\hat{s})
+ (4 m_A^2 m_{H_i}^2 - (\hat{t} + \hat{u})^2) (\hat{t} - \hat{u}) 
    C(m_{A}^2,m_{H_i}^2, \hat{s}) 
\nonumber \\ &\hspace*{5mm}
+(m_A^2 m_{H_i}^2 - \hat{t}^2)
\left[ T_A C(m_{A}^2,0,\hat{t}) + 
       T_i C(m_{H_i}^2,0,\hat{t}) 
\nonumber \right. \\ &\hspace*{35mm} \left.
+\hat{s} \hat{t}D(m_{A}^2,m_{H_i}^2,0,0,\hat{s},\hat{t})
\right]
\nonumber \\ &\hspace*{5mm}
-    (m_A^2 m_{H_i}^2 - \hat{u}^2) \left[ U_A 
     C(m_{A}^2,0,\hat{u}) + U_i C(m_{H_i}^2,0,\hat{u})
\nonumber \right. \\ &\hspace*{35mm} \left.
+\hat{s} \hat{u}D(m_{A}^2,m_{H_i}^2,0,0,\hat{s},\hat{u}) 
\right]
\nonumber \\ &\hspace*{5mm}
+ 2 m_q^2 (\hat{t} - \hat{u}) (m_A^2 m_{H_i}^2 - \hat{t} \hat{u})
\left[ D(m_{A}^2,0,m_{H_i}^2,0,\hat{t},\hat{u}) +
 D(m_{A}^2,m_{H_i}^2,0,0,\hat{s},\hat{t})
\nonumber \right. \\ &\hspace*{55mm} \left.
+ D(m_{A}^2,m_{H_i}^2,0,0,\hat{s},\hat{u})
\right] \Bigr\}  \;\; , 
\label{e10f}
\eeq \een
where $T_3^q$ is the third component of the weak isospin of the quark
running in the loop. 

\section{Numerical Results}

We are now ready to illustrate the importance of squark loop
contributions with a few examples.  For the numerical analysis we
have used the leading order CTEQ4L parameterization of the parton
distribution function of the proton \cite{cteq}, choosing the QCD
renormalization and factorization scales to be the sum of the
masses of the Higgs bosons in the final state. The effect of the
running mass of the bottom quark can be very important,
therefore, we have also included it in our calculations. In fact,
when the the main contribution to the processes comes from
bottom-quark loops and/or bottom-squark loops, the cross section
is proportional to the $\phi b \bar b$ Yukawa coupling to the
fourth power. Taking 3 and 5 GeV for typical running and pole
$b-$quarks mass respectively we can see that this effect can
reduce the Higgs pair production by a factor $(3/5)^4 \simeq
1/8$.

As discussed in Sec.~II.A.2, the squark loop diagrams shown in
Fig.~1 fall into three groups of diagrams, $(3)+(4)$, $ (5)+(6)$,
and $(7)+(8)$, the sum of diagrams in each group being finite and
gauge invariant.  In unpolarized $pp$ or $p \bar p$ scattering,
where only the sum of the square of all helicity amplitudes is
measurable, the squark contribution can therefore be
characterized by three loop functions and the associated products
of coupling constants.  In order to assess the importance of
these three sets of diagrams, we show in Fig.~3 their individual
contributions to the subprocess cross section ($\hat{\sigma}$)
for the production of $hh$ pairs. For the sake of simplicity,
only a single squark mass eigenstate ($\tilde{b}_1$) was included
here, whose mass is given on the $x-$axis.  We chose $m_h= 100$
GeV and $\sqrt{\hat{s}} = 3 m_h = 300$ GeV as typical values.
Note that diagrams (3), (4), (7) and (8) involve dimensionful
couplings, while in diagrams (5) and (6) only dimensionless
couplings appear. In order to show the corresponding loop
functions, we have therefore considered {\em fixed} ``typical''
coupling constants ($\text{$\tan \! \beta$}=50$,
$M_{\tilde{q}}=325$~GeV, $M_A=100$ GeV, $A_t=A_b=-410$~GeV,
$\mu=$-640~GeV).  However, these couplings were not varied as the
mass of the squark in the loop is changed, while physical
couplings do usually depend on the masses of the squark mass
eigenstates, {\em e.g.} through the change of the $\tilde{q}_L -
\tilde{q}_R$ mixing angle; see Appendix A.

The loop function describing diagrams $(3)+(4)$ is given in
Eqs.~(\ref{e7}a,b). It is the same (up to an overall factor) as
that describing the squark loop contribution to single Higgs
boson production \cite{gh2}. This contribution (dotted curve)
seems to be important only when real $H \to hh$ decays are
possible and \text{$\tan \! \beta$}\ is not too large, since
otherwise $Br(H \to hh)$ becomes very small. The contribution
from diagrams $(5)+(6)$ (dashed curve), which involves quartic
scalar couplings, is given in Eqs.~(\ref{e7}c,d). We find that
this contribution can increase the total cross section by no more
than a factor of a few. The reason is that this quartic scalar
coupling cannot significantly exceed the square of the
corresponding Yukawa coupling appearing in the quark loop
contribution, and the squarks in the loop cannot be much lighter
than the corresponding quarks. Of course, $m_{\tilde{b}_1} \gg
m_b$, but $m_{\tilde{t}_1} < m_t$ is still allowed.

The potentially largest contribution therefore comes from
diagrams $(7)+(8)$ (solid line), Eqs.~(\ref{e7}e--h), which
involve trilinear Higgs--squark--squark couplings. These
dimensionful couplings depend on unknown soft breaking
parameters, and might be (much) larger than the mass of the
lighter squark in the loop. On the other hand, while the loop
functions for diagrams $(3)+(4)$ and $(5)+(6)$ slightly increase
with increasing squark mass as long as $m_{\tilde q} <
\sqrt{\hat{s}}/2$, the loop function for diagrams $(7)+(8)$
starts to decrease as soon as $m_{\tilde q} > m_h/2$
\cite{foot:1}. Once $m_{\tilde q} > \sqrt{\hat{s}}/2$, all squark
loop functions become real, and drop rapidly with increasing
$m_{\tilde q}$, approximately like $m_{\tilde q}^{-4}$. Due to
the quick falling parton distribution functions, the largest
contribution to the total Higgs pair production cross section
come from values of $\hat{s}$ not far above threshold.  Figure 3
then shows that squark loop contributions to the total cross
section can only be large if the mass of the squarks in the loop
does not much exceed that of the produced Higgs bosons.

The results of Fig.~3 allowed us to search the MSSM parameter
space for parameters that maximize some subset of the diagrams
listed in Fig.~1. In these searches we imposed the following
constraints on parameters. First, most SUSY models predict
\be \label{e11}
1 < \text{$\tan \! \beta$} \leq \frac{ m_t(m_t) } {m_b(m_t)} \simeq 55 \;\; .
\ee
Second, we have interpreted the unsuccessful search for Higgs
bosons at LEP \cite{leph} to imply $m_h \geq 90$ GeV if the $Z Z
h$ coupling has similar strength as the corresponding coupling in
the SM. Otherwise the $Z A h$ coupling is large, and $(m_A + m_h)
\geq 175$ GeV is required. When computing the masses and
couplings of the CP--even Higgs bosons, we have included squark
and quark loop corrections \cite{yanagida} as given by the
1--loop effective potential \cite{ellis,dn,foot:2} . Turning to
the squark sector, for simplicity we took the same soft breaking
mass $m_{\tilde q}$ for $m_{\tilde{t}_L} \equiv m_{\tilde{b}_L},
\ m_{\tilde{t}_R}$ and $m_{\tilde{b}_R}$, and also took the same
value $A_q$ for the soft breaking parameters $A_t$ and $A_b$. The
squark sector is then completely determined by $m_{\tilde q}, \
A_q,\ \text{$\tan \! \beta$}$ and the supersymmetric Higgs mass
parameter $\mu$. In our scans we have imposed the LEP search
limit \cite{lepstop} $m_{\tilde{t}_1}, \ m_{\tilde{b}_1} \geq 80$
GeV \cite{foot:3}. 

As anticipated, we found that the potentially largest squark loop
contribution comes from diagrams $(7)+(8)$. In Figs.~4a and 4b we
show that (mostly) due to this contribution, squark loops can
increase the total $h-$pair cross section at the Tevatron by more
than two orders of magnitude if the mass of the lighter squark
eigenstate in the loop is close to its experimental lower limit.
In Fig.~4a we chose $m_A = 100$ GeV and $\text{$\tan \!
\beta$}=50$. The thick curves start at $(m_{\tilde q}, A_q, \mu)
= (325, -410, -640)$ GeV, corresponding to $h \tilde{b}_1
\tilde{b}_1$ coupling $V_{h \tilde{b}_1 \tilde{b}_1 } = 455$ GeV.
This choice saturates the LEP $Ah$ search limit (the $Z Z h$
coupling is very small here). The heavy solid and dashed curves
have been obtained by varying $m_{\tilde q}$ and $\mu$,
respectively, keeping all other parameters fixed. Note that
increasing $\mu$ ({\em i.e.} decreasing $|\mu|$) not only
increases $m_{\tilde{b}_1}$, but also reduces the $h \tilde{b}_1
\tilde{b}_1$ coupling, leading to a rapid drop--off of the squark
loop contribution. On the other hand, increasing $m_{\tilde q}$
leads to a somewhat slower decrease of this coupling. In both
cases the squark loop contribution to the total cross section
becomes essentially negligible for $m_{\tilde{b}_1} \geq 150$
GeV. For slightly smaller squark masses, there is mild
destructive interference between quark and squark loops. 

In Fig.~4b we have chosen $m_A = 500$ GeV and $\text{$\tan \!
\beta$} = 2.0$. The thick curves originate at $(m_{\tilde q},
A_q, \mu) = (380, 510, -975)$ GeV, which saturates the LEP $Z Z
h$ search limit \cite{foot:4}  and gives $V_{h \tilde{t}_1
\tilde{t}_1} = 475$ GeV. The thick heavy solid, dashed and dotted
lines have been obtained by varying, one at a time, $m_{\tilde
q}, \ A_q$ and $\mu$ respectively. We see that here the squark
loop contribution remains significant out to $m_{\tilde{t}_1}
\simeq 200$ GeV. This is partly due to the fact that the quark
loop contribution in Fig.~4b is about fifty times smaller than in
Fig.~4a, which in turn results from the large enhancement of the
$b-$loop contribution compared to the SM, by roughly a factor
$\tan^2 \beta =$ 2,500 in the amplitude. The bottom Yukawa
coupling in Fig.~4a is nearly as large as the top Yukawa coupling
in Fig.~4b. The former than gives a much larger quark loop
contribution than the latter, since for the relevant values of
$\hat{s}$, the absolute value of the (mostly imaginary) $b-$quark
loop function is much larger than that of the (mostly real)
$t-$quark loop function.

This enhancement of the contribution of $b-$quark loops at large
\text{$\tan \! \beta$}\ also implies that prospects for detecting
a signal for the production of neutral Higgs boson pairs at the
next Tevatron collider run might be better if \text{$\tan \!
\beta$}\ is large. We estimate that a cross section of 50 fb or
more might be detectable. This would lead to roughly 10 events
per experiment, each with 4 high$-p_T$ tagged $b-$jets and double
$b \bar{b}$ invariant mass peaking, assuming an integrated
luminosity of 2 fb$^{-1}$ and an overall efficiency of 10\%. The
enhancement of the pure quark loop contribution required to
achieve this cross section is given by the thin lines in Figs.~4a
and 4b. We see that for $\text{$\tan \! \beta$} = 50$ the total
cross section can exceed this sensitivity limit by more than an
order of magnitude. However, the maximal light squark mass
compatible with such a large $hh$ production cross section is
about the same at low and at large \text{$\tan \! \beta$}\
($\simeq 110$ GeV). 

It is interesting to notice that in Fig.~4a, $m_h$ is
significantly below $m_A$, {\em i.e.} the difference is up to 25
GeV, for all values of $m_{\tilde{b}_1}$ where squark loop
contributions are significant. This mass pattern, which is quite
unusual for large \text{$\tan \! \beta$}, is due to
non--logarithmic corrections to the Higgs boson mass matrix
involving trilinear scalar interactions; technically, due to
large contributions to $\Delta_{12}$, in the notation of
refs.~\cite{ellis,dn}. This means that the total cross sections
for the production of pairs of other Higgs bosons are quite small
at the Tevatron.

At high \text{$\tan \! \beta$}\ the Higgs pair production can be
enhanced by factors $\tan^4\beta$ or $1/\cos^4\beta$ with respect
to the SM production mechanism.  The only exception to this is
the $hh$ ($HH$) channel in the large (small) $M_A$ limit, in
which the factor $\sin\alpha / \cos\beta$ ($\cos\alpha /
\cos\beta$) appearing in Yukawa couplings goes to 1. For
instance, in the case that $(m_{\tilde q}, A_q, \mu) = (1, 1, 1)$
TeV, $\text{$\tan \! \beta$}=50$, and $M_A$=100 GeV the total
cross section at the Tevatron for the $hh$, $HH$, and $AA$
production is 3.3, 0.034, 3.9 fb respectively. When $M_A$ is
increased to 130 GeV these cross sections change to 0.15, 0.1,
and 0.7 fb.

The Higgs pair production cross sections do remain sizable at the
LHC. In Figs.~5a--d we show the squark loop contribution to $hh$
(a), $HH$ (b), $hH$ (c), and $AA$ (d) productions. The parameters
taken in Figs.~5a--c are the same as in Fig.~4a, which had been
chosen to maximize the $h \tilde{b}_1 \tilde{b}_1$ coupling. The
comparison of Figs.~4a and 5a shows that the relative importance
of squark loops is almost independent of the center--of--mass
energy. Of course, the total cross section increases greatly when
going from $\sqrt{s}=2$ TeV to 14 TeV due to the rapid increase
in the gluon--gluon luminosity. This is illustrated by the thin
lines, which again correspond to a total cross section of 50 fb.
The quark (mostly $b$) loop contribution by itself now exceeds
this cross section. Nevertheless, the background also increases
when going from the Tevatron to the LHC. In the absence of a
dedicated analysis of signal and background, we do not want to
claim that a total cross section of 50 fb necessarily gives a
detectable signal at the LHC, in spite of its considerably higher
anticipated luminosity.

Although the starting point of the curves in Figs.~5a--c had been
chosen to maximize $V_{h \tilde{b}_1 \tilde{b}_1}$, we find a
very large squark loop contribution also for $HH$ production
(Fig.~5b). In this case the squark loop contribution at first
{\em increases} with increasing $m_{\tilde{b}_1}$. The reason is
that the $H \tilde{b}_1 \tilde{b}_1 $ coupling increases quickly,
from $\sim 150$ GeV at the starting point of the curves to $\sim
350$ GeV near the maximum of the dark solid curve. In fact, this
coupling keeps increasing even further as $m_{\tilde q}$ is
increased, eventually reaching $\sim 450$ GeV. However, this
increase is overpowered by the rapid drop of the loop function
once $m_{\tilde{b}_1}$ significantly exceeds $m_H$ (see Fig.~3). 

The biggest relative contribution from squark loops appears in
$hH$ production, Fig.~5c, giving rise to an enhancement factor
$\sim 500$ in some cases. In this region of parameter space the $H
b \bar{b}$ Yukawa coupling becomes very small, due to the unusual
mixing pattern of CP--even Higgs bosons caused by radiative
corrections at large \text{$\tan \! \beta$}\ and large $|A_q|$
and $|\mu|$. At the peak of the curves the $H b \bar{b}$ coupling
vanishes completely, and the $h t \bar{t}$ coupling is quite
small, leading to a very small total quark loop contribution (see
the behavior of the thin lines). Notwithstanding, when $m_{\tilde
q}$ or $\mu$ are increased beyond this point, the quark loop
contribution reasserts itself while the squark loop contribution
decreases in absolute size, leading to a steep drop of the
relative importance of the squark contribution: it becomes
essentially negligible for $m_{\tilde{b}_1} \geq 220$ GeV. The
suppression of the $H b \bar{b}$ coupling also explains why
squark loop contributions to $HH$ production can remain
significant up to $m_{\tilde{b}_1} \simeq 300$ GeV (see Fig.~5b).
In this case the dominant quark contribution comes from top quark
loops, so squark loops remain significant unless
$m^2_{\tilde{b}_1} \gg m_t^2$.

For the parameters chosen in Figs.~5a--c the squark loop
contribution to $AA$ production is totally negligible. In this
case diagrams (7) and (8) in Fig.~1 have to include at least one
heavy squark mass eigenstate, since the diagonal $A \tilde{b}_1
\tilde{b}_1^*$ and $A \tilde{t}_1 \tilde{t}_1^*$ couplings vanish
identically. The off--diagonal $A \tilde{b}_1 \tilde{b}_2^*$
coupling is actually quite large, $\sim 270$ GeV at the starting
point of the curves in Fig.~5a. However, at the same time
$m_{\tilde{b}_2} = 455$ GeV, which suppresses the contributions
from diagrams $(7)+(8)$ to an insignificant level.

In Fig.~5d, we therefore show results for a scenario with
relatively small $m_{\tilde{t}_2}$: $m_A = 150$ GeV, $\text{$\tan
\! \beta$} = 4, \ A_q = -110$ GeV, $\mu = 345$ GeV, and
$m_{\tilde q}$ between 115 and 350 GeV. The same set of
parameters also yields a relatively large quartic $A A
\tilde{t}_1 \tilde{t}_1^*$ coupling, so that diagrams $(5)+(6)$
in Fig.~1 are maximized. Since \text{$\tan \! \beta$}\ is fairly
small, the quark loop contribution is dominated by top quark
loops; however, the loop function is also different for CP--odd
Higgs bosons, see Eqs.(6), and  its contribution is suppressed by
a factor $\cot^4 \beta \simeq 1/250$ compared to the SM case.
Therefore diagrams $(7)+(8)$  can make a large relative
contribution, as long as $m_{\tilde q} \leq m_t$. Even
though squark loop contributions can increase the $AA$ cross
section by a factor of about 200 in this case, it still remains
well below the cross section for $HH$ production. In scenarios
where the $AA$ cross section is comparable to the $HH$ cross
section, we find that squark loop corrections to $AA$ production
are quite modest.

Given that we are only working in leading order in QCD, our
predictions for the total cross sections have significant
uncertainties due to the choice of scale in $\alpha_s$, $m_b$,
and the parton distribution functions. We took the same scale
everywhere, {\em viz.} the sum of the masses of the produced
Higgs bosons. One could therefore infer the existence of squark
loop contributions to the total cross section only if it changes
the quark loop result by at least a factor of two \cite{foot:5}.
However, smaller squark loop contributions might still be visible
in some distributions. For example, for large \text{$\tan \!
\beta$}\ the quark loop contribution is dominated by $b-$quark
loops (except for $hh$ production at large $m_A$, where $h$ is
always SM--like). Since $m_{\tilde{b}_1} \gg m_b$, the squark and
quark loop functions show quite a different dependence on
$\hat{s}$. This is illustrated in Fig.~6, where we show $d
\sigma(hh) /d \sqrt{\hat{s}}$ for the point in Fig.~5a where $b$
and $\tilde{b}$ loops contribute equally. The solid histogram
shows the contribution of $b$ loops only, while the dashed
histogram includes squark loops; the total cross section (area
under the curves) has been normalized to be the same in both
cases. The $b$ loop contribution peaks just beyond the threshold
at $\sqrt{\hat s} = 2 m_h$, but the $\tilde{b}_1$ loop
contribution clearly shows up as a second peak at $\sqrt{\hat{s}}
\simeq 2 m_{\tilde{b}_1}$. This distribution is in principle
directly measurable if both Higgs bosons decay hadronically (with
combined branching ratio $\simeq 80\%$). Given sufficient
statistics, one might be able to infer the squark loop
contribution in this manner even if the total cross section is
dominated by quark loops.

\section{Summary and Conclusions}

In this paper we have calculated squark loop contributions to the
pair production of two neutral Higgs bosons. If CP is conserved,
squark loops contribute only if the two produced Higgs bosons
have identical CP quantum numbers. In Sec.~II we gave complete
analytical expressions that allow the evaluation of these
contributions. For completeness we also included expressions for
the quark loop contributions \cite{plehn}. These formulae are
written in a completely general fashion with the explicit
expressions for the relevant coupling constants in the framework
of the MSSM being listed in Appendix A.

In Sec.~III we showed some numerical results for the MSSM. We found
that squark loop contributions can increase the total cross
section for the production of two CP--even Higgs bosons by more
than two orders of magnitude. However, such large contributions
are possible only if three conditions are satisfied:
\begin{itemize}
\item The relevant Yukawa coupling must be large. In case of the
(s)top, this is always true for at least one of the two CP--even
Higgs bosons of the MSSM. However, as pointed out quite some time
ago \cite{barnett}, in the MSSM (as in other models with more
than one Higgs doublet) the bottom Yukawa coupling can also be
large, if the vacuum expectation value that gives rise to the
mass of the $b$ quark is small. In the MSSM this happens for
large values of \text{$\tan \! \beta$}.  However, the bottom
Yukawa coupling is not expected to exceed that of the top quark.

\item The lighter of the two superpartner whose Yukawa coupling
is large (generally $\tilde{b}_1$ at large \text{$\tan \!
\beta$}\ and $\tilde{t}_1$ at small \text{$\tan \! \beta$}) must
not be much heavier than the Higgs bosons in the final state.
This condition is especially critical at large \text{$\tan \!
\beta$}, since here $\tilde{b}_1$ loops have to compete with $b$
quark loops. For equal Yukawa couplings ({\em i.e.} $\text{$\tan
\! \beta$} \simeq m_t/m_b$), the squared $b-$loop contribution
exceeds the squared $t-$loop contribution by a factor of $\sim
50$. If this condition is satisfied, open squark pair production
should be detectable at the same collider, unless the squark--LSP
mass difference is very small.

\item The relevant trilinear soft breaking parameters and/or
$|\mu|$ must be significantly larger than the mass of the lighter
squark eigenstate. This implies that the diagonal entries of the
corresponding squark mass matrix must also significantly exceed
the smaller eigenvalue of this matrix. This requires a modest
amount of fine tuning. However, in the absence of a complete
theory of Supersymmetry breaking this possibility should not be
discounted.
\end{itemize}

If these three conditions are satisfied, the region $m_A \simeq
100$ GeV can perhaps even be probed at the next run of the
Tevatron collider, using searches for final states with 4
high$-p_T \ b-$jets. Moreover, this process has a big advantage
over the $\phi b\bar{b}$ associated production \cite{yuan}, which
leads to the same final state: the reconstruction efficiency of
Higgs pairs is expected to be one order higher than that for
$\phi  b \bar{b}$ production. This originates from the fact that
{\em all} $b$-jets coming from Higgs boson pair decays are
energetic, while the two associated $b$-jets in the $\phi b
\bar{b}$ channel are quite soft.  At the same time the process
under study has a cross section not much smaller than that for
$\phi b \bar{b}$. The reach of the LHC should be much higher, but
a quantitative statement will only be possible after a detailed
analysis of signal and background \cite{future}. Furthermore, we
should point out that our result for the 1--loop cross sections
are probably very conservative, since we can expect large QCD
corrections, which could increase the cross section by as much as
a factor of $\sim 2$ \cite{nlo1,nlo2,nlo3}. 

Squark loop contributions to the pair production of two CP--odd
Higgs bosons are more modest in general. In this case trilinear
$A \tilde{q}_i \tilde{q}^*_j$ couplings contribute only if at
least one squark in the loop is a heavy mass eigenstate, which
leads to a suppression of the squark contributions. Nevertheless,
squark loops can give rise to large enhancements of the $A$ pair
production cross section if $m_{\tilde{t}_1} \le m_t$ and
$\text{$\tan \! \beta$} \sim 5$, which leads to a very small
quark loop contribution. However, the total cross section for
$AA$ production remains quite small in this case.

In general $AA$ final states, as well as the $hA$ and $HA$ final
states, which receive no contributions from squark loops (but do
receive Drell-Yan like contributions from light $q \bar{q}$
annihilation \cite{plehn}), can be significant. Note that in most
regions of parameter space the CP--odd Higgs boson $A$ is nearly
degenerate with one of the two CP--even Higgs bosons. This means
that often three different channels ({\em e.g.} $HH, \ HA$ and
$AA$) contribute to essentially the same final state, and
therefore have to be added. This obviously increases the chance
to detect a signal for Higgs pair production in the $4b$ channel.
At the same time it complicates the interpretation of such a
signal, {\em e.g.} the extraction of the relevant coupling
constants. Nevertheless, we saw in Fig.~6 that the analysis of
the $4b$ invariant mass distribution can help to disentangle the
various contributions to the signal. We are therefore hopeful
that the search for the pair production of neutral Higgs bosons
in the $4b$ channel will provide information that will help us to
pin down the Higgs sector, and perhaps also the squark sector of
the theory.

\acknowledgments

A.B., M.D., and J.K.M. were supported by FAPESP, Brazil. A.B.
would like to thank A.\ Solomin for the help with the computing
facilities. This work was supported by Conselho Nacional de
Desenvolvimento Cient\'{\i}fico e Tecnol\'ogico (CNPq), by
Funda\c{c}\~ao de Amparo \`a Pesquisa do Estado de S\~ao Paulo
(FAPESP), by Programa de Apoio a N\'ucleos de Excel\^encia
(PRONEX) and by  U.S. Department
of Energy under the contract DE-AC03-76SF00515.

\appendix
\section{MSSM Coupling Constants}

We denote the weak mixing angle and couplings as
\be
s_W\equiv \sin\theta_W \;\; , \;\;  
c_W\equiv\cos\theta_W  \;\; , \;\;  
g=e/s_W \;\; , \;\;  
g_Z=g/c_W \;\; .
\ee

We define squark mass eigenstates via
\be \label{ea1}
    \llgm{\tilde f_1 \cr \tilde f_2} \rrgm =
    \llgm{\cos\theta_f & \sin\theta_f \cr -\sin\theta_f & \cos\theta_f}\rrgm
    \llgm{\tilde f_L \cr \tilde f_R}\rrgm,~~~~f=u,d \;\; .
\ee
where the mixing angle $\theta_f$ is defined in such way that the mass
matrix becomes diagonal with eigenvalues $m^2_{\tilde f_1}$ and
$m^2_{\tilde f_2}$ ($m_{\tilde f_1}<m_{\tilde f_2}$):
\be \label{ea2}
  \left(\begin{array}{l l}
    \cos\theta_f & \sin\theta_f \\ 
   -\sin\theta_f & \cos\theta_f
  \end{array} \right)
  \left(\begin{array}{l l}
     m^2_{\tilde f_L} & m^2_{\tilde f_{LR}} \\ 
     m^{2*}_{\tilde f_{LR}} & m^2_{\tilde f_R} 
  \end{array}\right)
  \left(\begin{array}{l l}
    \cos\theta_f & \sin\theta_f \\ 
   -\sin\theta_f & \cos\theta_f
  \end{array} \right) \\
  =
  \left(\begin{array}{l l}
          m^2_{\tilde f_1} & 0 \\
          0 & m^2_{\tilde f_2} \\
  \end{array} \right) 
\ee
with
\ben \label{ea3} \beq
    m^2_{\tilde f_L} =& \widetilde m^2_{\tilde f_L}+ m^2_f 
                        + M_Z^2\cos 2\beta(T_{3}^f - Q^f s^2_W) \;\; ;
\label{ea3a} \\
    m^2_{\tilde f_R} =&~ \widetilde m^2_{\tilde f_R}+ m^2_f 
                        + M_Z^2\cos2\beta Q_fs^2_W \;\; ;
\label{ea3b} \\
    m^2_{\tilde f_{LR}}=& 
    \left\{ \begin{array}{l l}
    -m_u(\mu \cot\beta-A_u^*), & \ \ f=u\\
    -m_f(\mu \tan\beta-A_f^*), & \ \ f=d
    \end{array} 
    \right. \;\; .
\label{ea3c}
\eeq \een

Squark current eigenstate bilinears can then be expressed in terms of
current mass as follows:
\ben \label{ea4} \beq
\tilde{f_L^*}\tilde{f_L} &=
 +c_f^2   \tilde{f_1^*}\tilde{f_1}
 -c_f s_f \tilde{f_1^*}\tilde{f_2}
 -c_f s_f \tilde{f_2^*}\tilde{f_1}
 +s_f^2   \tilde{f_2^*}\tilde{f_2} ;
\label{ea4a} \\
\tilde{f_L^*}\tilde{f_R} &=
+c_f s_f \tilde{f_1^*}\tilde{f_1}
+c_f^2   \tilde{f_1^*}\tilde{f_2}
-s_f^2   \tilde{f_2^*}\tilde{f_1}
-c_f s_f \tilde{f_2^*}\tilde{f_2} ;
\label{ea4b} \\
\tilde{f_R^*}\tilde{f_L} &=
+c_f s_f \tilde{f_1^*}\tilde{f_1}
-s_f^2   \tilde{f_1^*}\tilde{f_2}
+c_f^2   \tilde{f_2^*}\tilde{f_1}
-c_f s_f \tilde{f_2^*}\tilde{f_2} ;
\label{ea4c} \\
\tilde{f_R^*}\tilde{f_R} &=
+s_f^2   \tilde{f_1^*}\tilde{f_1}
+c_f s_f \tilde{f_1^*}\tilde{f_2}
+c_f s_f \tilde{f_2^*}\tilde{f_1}
+c_f^2   \tilde{f_2^*}\tilde{f_2} ,
\label{ea4d}
\eeq \een
with $c_f=\cos \theta_f$ and $s_f=\sin\theta_f$ for $f=u,d$.

After these preliminaries we are ready to list the relevant
couplings. We list couplings to squark current eigenstates only; these
can be converted into couplings to mass eigenstates using Eqs.~(17).
   
\vspace*{3mm}
\noindent{\bf H(h, A)-Fermion-Fermion}\par

\medskip

\begin{tabular}{l l l l l l}
$\bar u u h^0 :$&$   -{{gm_u\cos\alpha}\over{2M_W\sin\beta}} $&\ \ \ 
$\bar u u H^0 :$&$   -{{gm_u\sin\alpha}\over{2M_W\sin\beta}} $&\ \ \ 
$\bar u u A^0 :$&$   +i{{gm_u}\over{2M_W}}\cot\beta \gamma_5 $\\
$\bar d d h^0 :$&$   +{{gm_d\sin\alpha}\over{2M_W\cos\beta}} $&\ \ \
$\bar d d H^0 :$&$   -{{gm_d\cos\alpha}\over{2M_W\cos\beta}} $&\ \ \
$\bar d d A^0 :$&$   +i{{gm_d}\over{2M_W}}\tan\beta \gamma_5 $
\end{tabular}

\vspace*{3mm}
\noindent{\bf Gluon--Squark--Squark}\par

\medskip

\begin{tabular}{l l }
$G^{\mu\alpha}\tilde f^*_i(p) \tilde f_i(k):  $&\ \ \ \
$-ig_s {{\lambda^\alpha}\over 2} (k-p)^\mu$
\end{tabular}
\\
where $k$ and $p$ are the incoming momenta of the squarks
$\tilde{q}_i$, with $i=1,2$.

\vspace*{3mm}
\noindent{\bf H-Squark-Squark}\par

\medskip

\begin{tabular}{l l }
$\tilde u^*_L\tilde u_L H^0: \ \ \ $ & $
 -({{gm_u^2\sin\alpha}\over{M_W\sin\beta}} + g_ZM_Z\cos(\alpha+\beta)
 ({1\over 2}-{2\over3}s_W^2))     $\\
$\tilde u^*_R\tilde u_R H^0: \ \ \ $&$
 -({{gm_u^2\sin\alpha}\over{M_W\sin\beta}} +{2\over 3}g_ZM_Z
   \cos(\alpha+\beta) s_W^2)$\\
$\tilde u^*_L\tilde u_R H^0,\tilde u^*_R\tilde u_L H^0: \ \ \ $&$
 +{{gm_u}\over{2M_W\sin\beta}}(A_u\sin\alpha + \mu\cos\alpha) $\\
$\tilde d^*_L\tilde d_L H^0: $ & $  
 -({{gm_d^2\cos\alpha}\over{M_W\cos\beta}} + g_ZM_Z\cos(\alpha+\beta)
       (-{1\over 2}+{1\over3}s_W^2)) $ \\
$\tilde d^*_R\tilde d_R H^0: $&$
 -({{gm_d^2\cos\alpha}\over{M_W\cos\beta}} -{1\over 3}g_ZM_Z
           \cos(\alpha+\beta)s_W^2) $\\
$\tilde d^*_L\tilde d_R H^0,\tilde d^*_R\tilde d_L H^0:  $&$
 +{{gm_d}\over{2M_W\cos\beta}}(A_d\cos\alpha + \mu\sin\alpha)$
\end{tabular}

\vspace*{3mm}
\noindent{\bf h-Squark-Squark}\par

\medskip

$
 h\tilde f^*\tilde f = H\tilde f^*\tilde f
\{  \sin\alpha \to \cos\alpha,\cos\alpha \to -\sin\alpha, 
 \sin(\alpha+\beta) \to \cos(\alpha+\beta), 
 \\
\hspace*{33mm} \cos(\alpha+\beta) \to -\sin(\alpha+\beta)\} $

\vspace*{3mm}
\noindent{\bf A-Squark-Squark}\par

\medskip

\begin{tabular}{l l l l}
$      \tilde u_L^*\tilde u_R A^0: $&
$\  -i{{gm_u}\over {2M_W}}(A_u\cot\beta-\mu) $&$
\ \ \  \tilde u_R^*\tilde u_L A^0: $&
$\  +i{{gm_u}\over {2M_W}}(A_u\cot\beta-\mu) $\\
$\tilde d_L^*\tilde d_R A^0: $&
$\  -i{{gm_d}\over {2M_W}}(A_d\tan\beta-\mu) $&$
\ \ \  \tilde d_R^*\tilde d_L A^0: $&
$\ +i{{gm_d}\over {2M_W}}(A_d\tan\beta-\mu) $ 
\end{tabular}

\vspace*{3mm}
\noindent{\bf Gluon-Gluon-Squark-Squark}\par

\medskip

$g_\mu^\alpha g^{\mu\beta}\tilde q^*_i\tilde q_i: 
\ \ \ g_s^2 {{\lambda^\alpha}\over 2} {{\lambda^\beta}\over 2}
$

\vspace*{3mm}
\noindent{\bf H-H-Squark-Squark}\par

\medskip

\begin{tabular}{l l}
$\tilde u_L^*\tilde u_L (H^0)^2:$ & $
 - [{{g^2m_u^2\sin^2\alpha}\over{4M_W^2 \sin^2\beta}}
       +{1\over 4}g_Z^2(T_{3u}-s_W^2Q_u)\cos 2\alpha]$ \\ 
$ \tilde u_R^*\tilde u_R (H^0)^2:$ & $
 - [{{g^2m_u^2\sin^2\alpha}\over{4M_W^2 \sin^2\beta}} 
         +{1\over 4}g_Z^2s_W^2Q_u\cos 2\alpha] $\\ 
$\tilde d_L^*\tilde d_L (H^0)^2: $ & $ 
- [{{g^2m_d^2\cos^2\alpha}\over{4M_W^2 \cos^2\beta}}
     +{1\over 4}g_Z^2(T_{3d}-s_W^2Q_d)\cos 2\alpha] $ \\
$\tilde d_R^*\tilde d_R (H^0)^2:$&$
- [{{g^2m_d^2\cos^2\alpha}\over{4M_W^2 \cos^2\beta}}
                  +{1\over4 }g_Z^2s_W^2Q_d\cos 2\alpha]$
\end{tabular}

\vspace*{3mm}
\noindent{\bf h-h-Squark-Squark}\par

\medskip

$hh\tilde q\tilde q=HH\tilde q \tilde q
\{\sin\alpha \to -\cos\alpha,
\cos\alpha \to \sin\alpha,
\cos 2\alpha \to -\cos 2\alpha \}
$

\vspace*{3mm}
\noindent{\bf H-h-Squark-Squark}\par

\medskip

$Hh\tilde q\tilde q=HH\tilde q \tilde q
      \{\cos 2\alpha \to -\sin 2\alpha,\sin^2\alpha \to \sin 2\alpha,
        \cos^2\alpha \to -\sin 2\alpha\}
$

\vspace*{3mm}
\noindent{\bf A-A-Squark-Squark}\par

\medskip

$AA\tilde q\tilde q=HH\tilde q \tilde q
     \{\sin\alpha \to -\cos\beta,
       \cos\alpha \to \sin\beta,
       \cos 2\alpha \to -\cos 2\alpha\}
$

\vspace*{3mm}
\noindent{\bf Higgs$^0$-Higgs$^0$-Higgs$^0$}\par

\medskip

\begin{tabular}{l l l l}
$(H^0)^3:    $&$ -{{g_Z}\over 4}M_Z \cos 2\alpha\cos(\alpha+\beta)$&$
\ \ \ (h^0)^3:     $&$ -{{g_Z}\over 4}M_Z \cos 2\alpha\sin(\alpha+\beta)$\\
$(A^0)^2 H^0:$&$ +{{g_Z}\over 4}M_Z \cos 2\beta\cos(\alpha+\beta)$&$
\ \ \ (A^0)^2 h^0: $&$ -{{g_Z}\over 4}M_Z \cos 2\beta\sin(\alpha+\beta)$
\end{tabular}

\begin{tabular}{l l}
$(h^0)^2 H^0:$&$ -{{g_Z}\over 4}M_Z[2\sin 2\alpha\sin(\alpha+\beta)
                                  - \cos 2\alpha\cos(\alpha+\beta)]$\\
$(H^0)^2 h^0:$&$ +{{g_Z}\over 4}M_Z[2\sin 2\alpha\cos(\alpha+\beta)
                                  + \cos 2\alpha\sin(\alpha+\beta)]$
\end{tabular}

\vspace*{3mm}
\noindent{\bf Z-Higgs-Higgs}\par

\medskip

\begin{tabular}{l l l l}
$h^0(p)A^0(k)Z_\mu :$ &$ +{ig_z\over{2}}\cos(\alpha-\beta)(k-p)^\mu $&  \ \ \ \
$h^0(p)A^0(k)Z_\mu :$ &$ +{ig_z\over{2}}\sin(\alpha-\beta)(k-p)^\mu $
\end{tabular}

\label{coup}

\section{Conventions}
\label{int}

We chose the following polarization vectors in our calculations
\[
\epsilon_1^+ = \epsilon_2^- = \frac{1}{\sqrt{2}} (0, -1, -i, 0) \;\; ,
\]
\[
\epsilon_1^- = \epsilon_2^+ = \frac{1}{\sqrt{2}} (0,  1, -i, 0) \;\; ,
\]
where the first gluon is moving along the $z$ axis.

We have  used scalar $C_0$ and $D_0$ Passarino-Veltman functions
in our analytical formulas which are expressed as:
\ben \beq
&C_0[p10,p12,p20,m0,m1,m2]=\nonumber\\
&(i\pi^2)^{-1}\int d^4q
  ([q^2-m_0^2][(q+p_1)^2-m_1^2][(q+p_2)^2-m_2])^{-1}
\;\; , \\
&D_0[p10,p12,p23,p30,p20,p13,m0,m1,m2,m3]=\nonumber \\
&(i\pi^2)^{-1}\int d^4q
([q^2-m_0^2][(q+p_1)^2-m_1^2][(q+p_2)^2-m_2][(q+p_3)^2-m_3^2])^{-1} 
\;\; .
\eeq \een
Our  convention for the scalar arguments is
$pi0=p_i^2$,  $pij=(p_i-p_j)^2$ and $mi=m_i$.


\newpage

\begin{figure}
\mbox{\epsfig{file=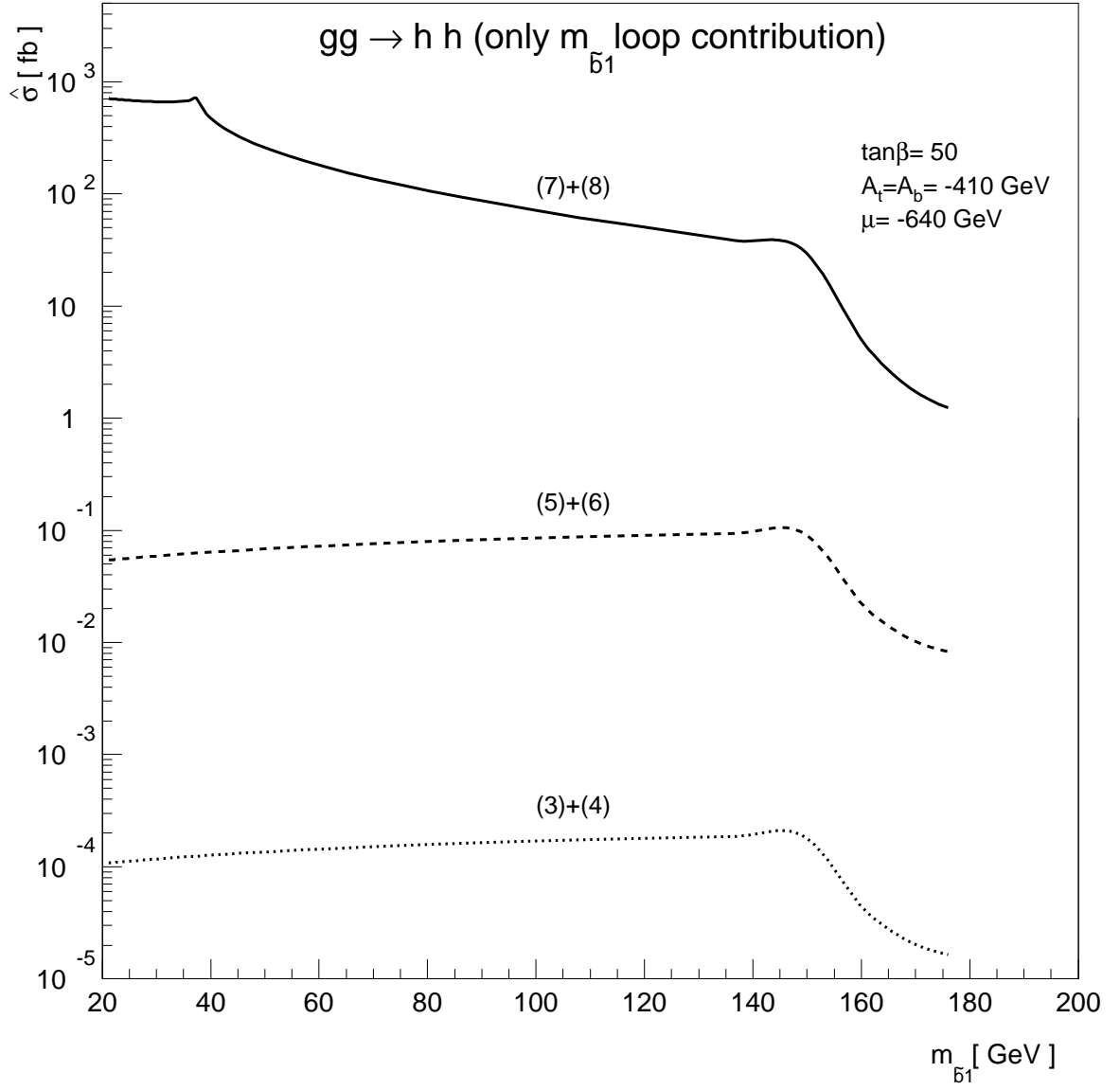,width=\textwidth}}
\caption{Contributions of the $(3)+(4)$, $(5)+(6)$, and $(7)+(8)$
diagram sets to the subprocess cross section for the production of
$hh$ pairs as a function of $m_{\tilde{b}_1}$. 
We included only the $\tilde{b}_1$ effects and assumed $m_h =
100$ GeV and $\protect\sqrt{\hat{s}}= 300$ GeV. }
\label{fig:test}
\end{figure}


\begin{figure}
\mbox{\epsfig{file=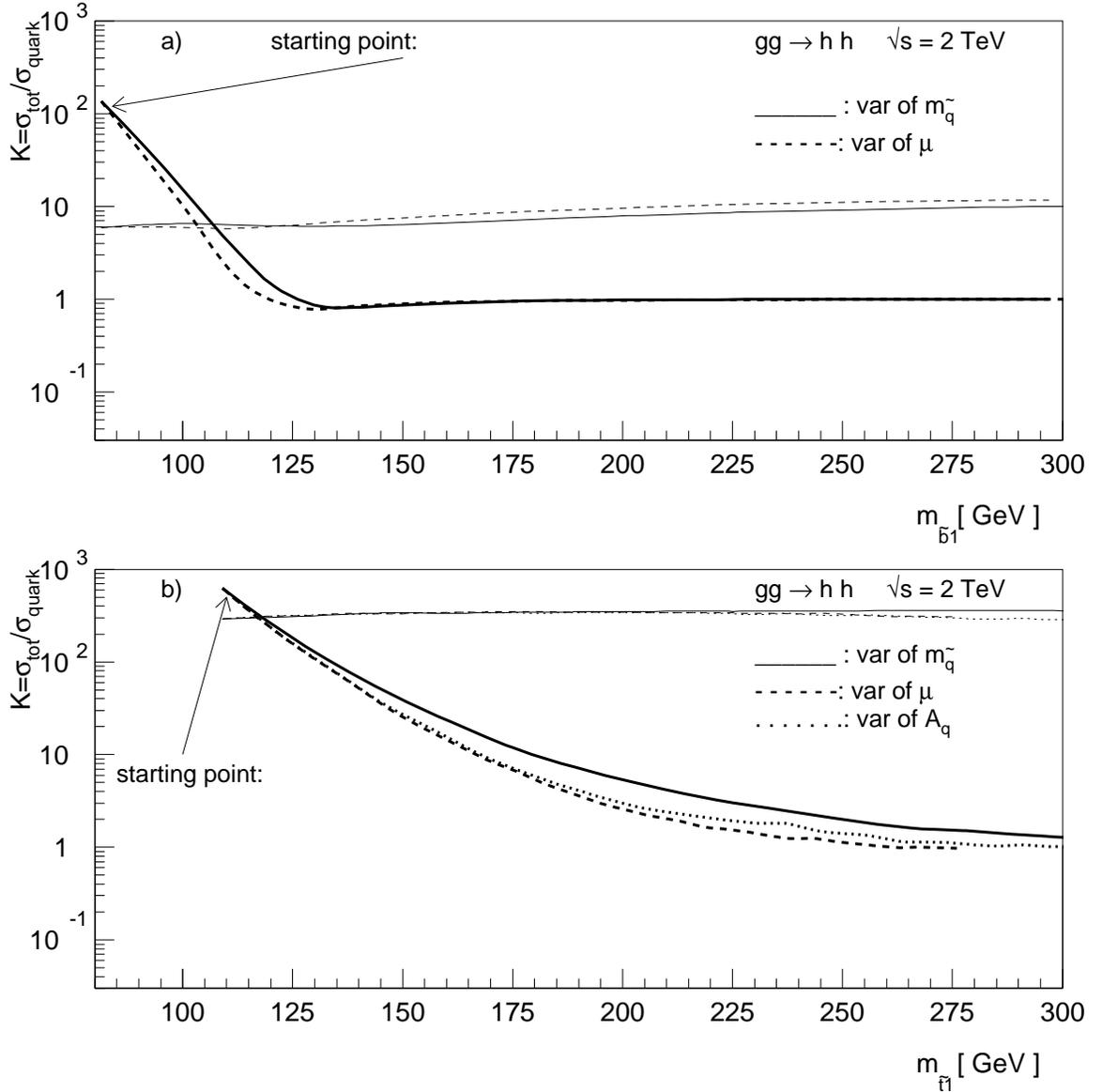,width=\textwidth}}
\caption{Ratio of the total cross section to the cross section including
only the quark contributions for $gg \to hh$ at the Tevatron. (a)
corresponds to a large \text{$\tan \! \beta$}\ $(=50)$ scenario and we took $(m_{\tilde
q}, A_q, \mu) = (325, -410, -640)$ GeV. The heavy solid and dashed
curves have been obtained by varying, one at a time, $m_{\tilde q}$
and $\mu$, respectively. In (b) we display a low \text{$\tan \! \beta$}\ ($=2$)
scenario using $(m_{\tilde q}, A_q, \mu) = (380, 510, -975)$ GeV and
conventions as in (a). Here, the heavy dotted line was obtained by
varying $A_q$. In both (a) and (b), the thin lines correspond to the
enhancement needed for the total cross section to be at the level of 50
fb.}
\label{fig:hh_1}
\end{figure}


\begin{figure}
\mbox{\epsfig{file=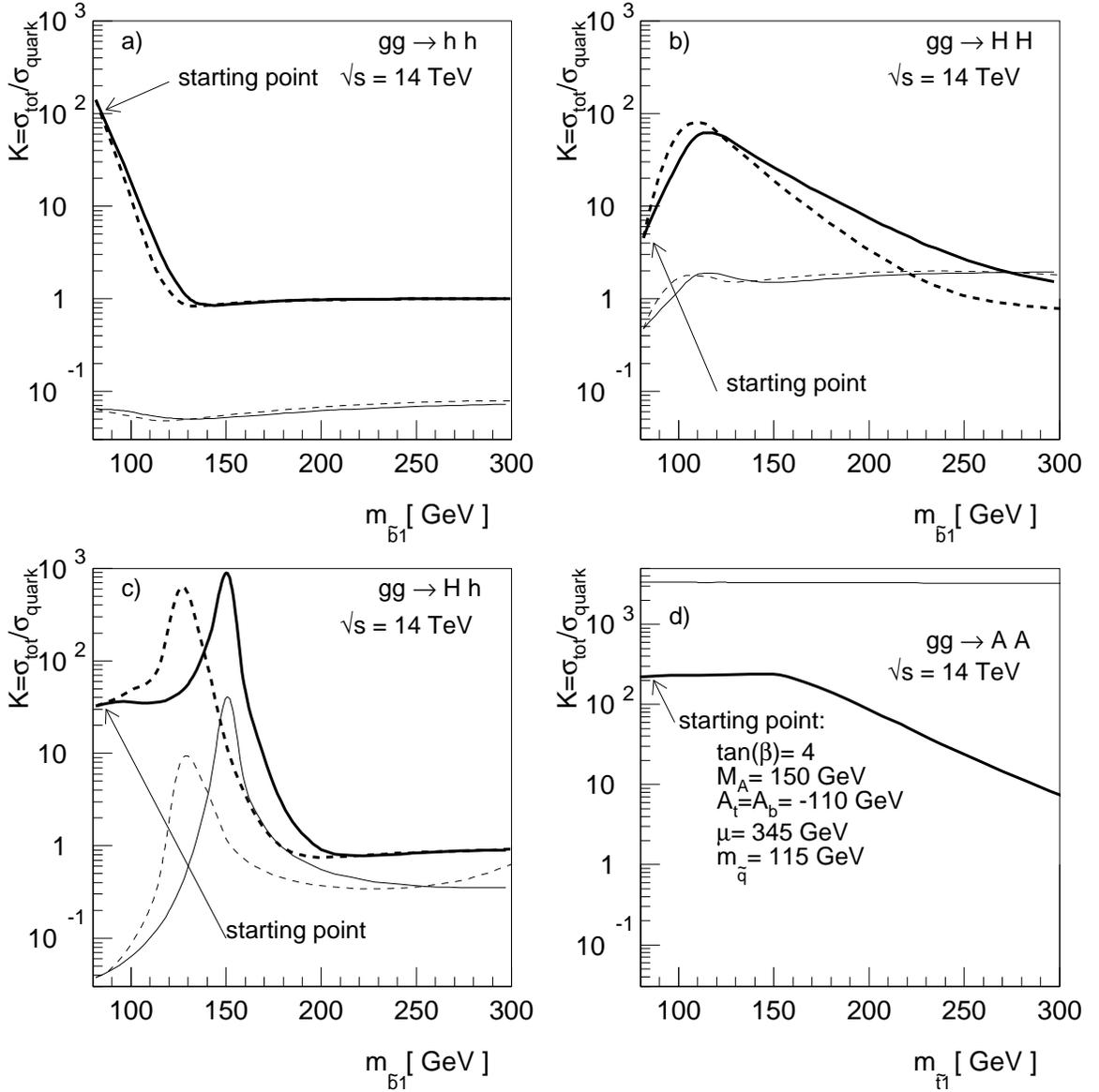,width=\textwidth}}
\caption{Same as in Fig.~\protect\ref{fig:hh_1} for the $hh$ (a),
$HH$ (b), $Hh$ (c) and $AA$ (d) production at the LHC. The
conventions are the same as in Fig.~\protect\ref{fig:hh_1}. In
(a)--(c) we chose the parameters used for Fig.~4a. In (d) we
assumed that $m_A = 150$ GeV, $\text{$\tan \! \beta$} = 4$, $A_q
= -110$ GeV, $\mu = 345$ GeV, and $m_{\tilde q}$ between 115 and
350 GeV.}
\label{fig:hh_2}
\end{figure}


\begin{figure}
\mbox{\epsfig{file=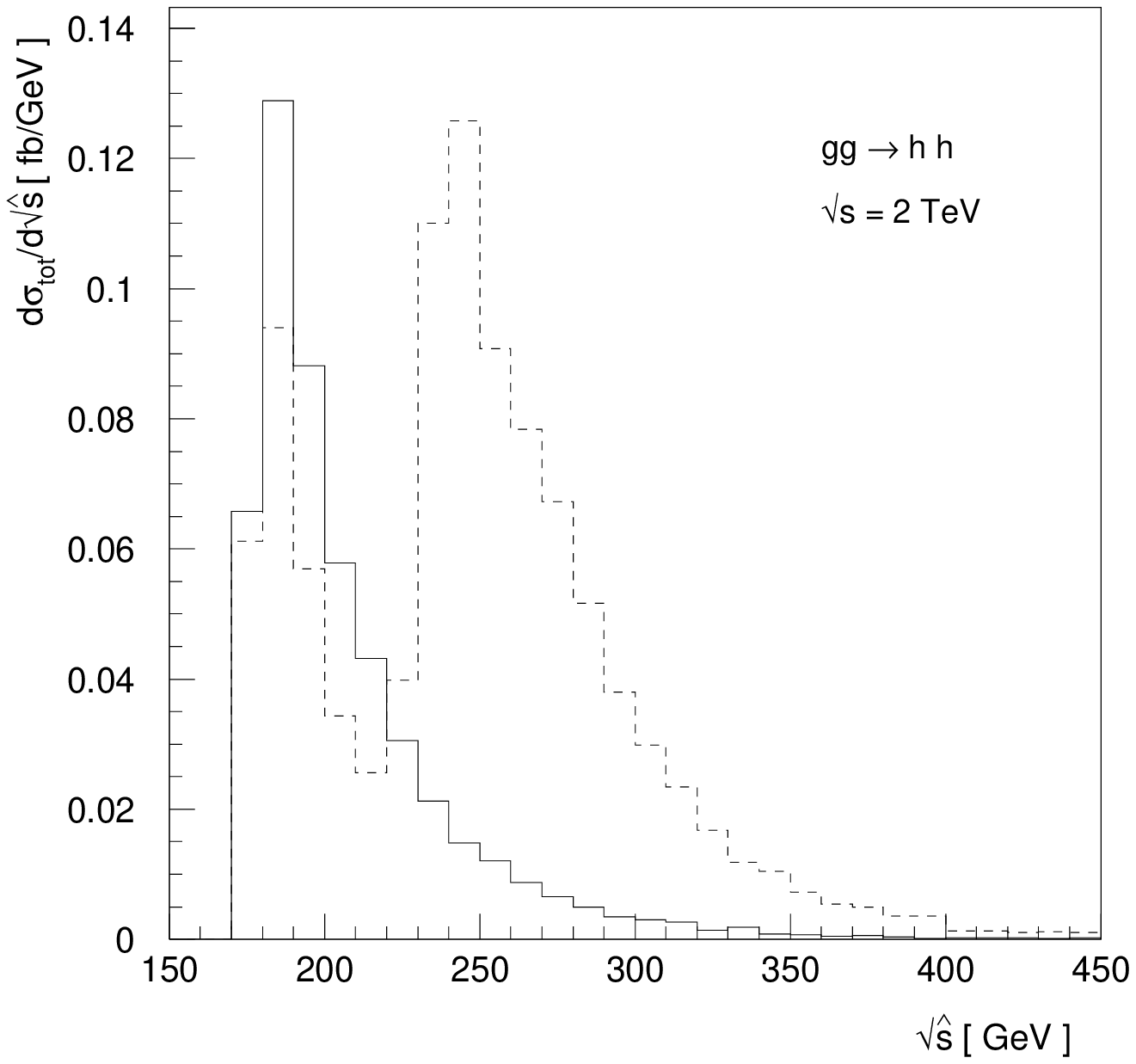,width=\textwidth}}

\caption{$d\sigma/d\protect{\sqrt{\hat{s}}}$ including only quark
loops (solid line) and considering quark and squark loops (dashed
line).  We have chosen $(m_{\protect{\tilde{q}}},A_q,\mu)=(335,-410-640)$, 
$M_A=100$ GeV, and $\text{$\tan \! \beta$}=50$ for which quark and squark loops
contribute almost equally: $\sigma_{total}=10.$ fb,
$\sigma_{squark}=5.6$ fb.}
\label{fig:shat}
\end{figure}


\begin{references}
\bibitem{foot:1} Of course, squark masses below 50 GeV are
excluded experimentally. We extend the curves in Fig.~3 to such
small values of $m_{\tilde q}$ since the shapes of these curves
remain the same when $m_{\tilde q}, \ m_h$ and $\sqrt{\hat{s}}$
are all scaled up by the same factor. In some regions of
parameter space $H \to \tilde{t}_1 \tilde{t}_1^*$ or $H \to
\tilde{b}_1 \tilde{b}_1^*$ decays are still possible
\cite{bartl}.
\bibitem{foot:2} The expressions in \cite{carena} that include RG
improvement and leading 2--loop corrections are not applicable
for large splitting between the squark masses; such cases are
important for us. Note also that much of these effects can be
incorporated into the 1--loop expression \cite{dn,haber} by using
running quark masses at the appropriate scale.
\bibitem{foot:3} Searches at the Tevatron \cite{stoptev} for a single
squark mass eigenstate decaying into a stable neutralino plus a
light quark exclude squark masses up to $\sim 120$ GeV, but only
if the neutralino mass is less than $\sim 50$ GeV. Under the same
assumptions the LEP search limit remains valid as long as the
squark--neutralino mass splitting exceeds $\sim 5$ GeV.
\bibitem{foot:4} Preliminary results from the LEP run at
$\sqrt{s}=189$ GeV \cite{pepe} indicate $m_{h} \geq 95$ GeV for
SM--like $h$. This would exclude only the very top of the curves
in Fig.~4b, reducing the maximal squark loop contribution in this
case by about a factor of 2.
\bibitem{foot:5} We are assuming here that the relevant Yukawa
couplings, {\em i.e.} the Higgs mixing angles $\alpha$ and
$\beta$, are known from other measurements; otherwise the
uncertainty in the quark loop contribution is much larger.
\end{references}
\end{document}